\documentclass[aps,showpacs,twocolumn,pra]{revtex4}

\usepackage[tbtags]{amsmath}
\usepackage{amssymb}
\usepackage{amsfonts}
\usepackage[dvips]{graphicx,color}

\begin{document}
\title{Clock synchronization using maximal multipartite entanglement}
\author{Changliang Ren}
\author{Holger F. Hofmann}
\email{hofmann@hiroshima-u.ac.jp} \affiliation{ Graduate School of
Advanced Sciences of Matter, Hiroshima University, Kagamiyama 1-3-1,
Higashi Hiroshima 739-8530, Japan} \affiliation{JST, CREST,
Sanbancho 5, Chiyoda-ku, Tokyo 102-0075, Japan}

\begin{abstract}
We propose a multi party quantum clock synchronization protocol that
makes optimal use of the maximal multipartite entanglement of
GHZ-type states. To realize the protocol, different versions of
maximally entangled eigenstates of collective energy are generated
by local transformations that distinguish between different
groupings of the parties. The maximal sensitivity of the entangled
states to time differences between the local clocks can then be
accessed if all parties share the results of their local time
dependent measurements. The efficiency of the protocol is evaluated
in terms of the statistical errors in the estimation of time
differences and the performance of the protocol is compared to
alternative protocols previously proposed.
\end{abstract}


\maketitle

\section{Introduction}
Quantum clock synchronization protocols are of fundamental interest in quantum information, since they can illustrate how information about time is encoded in quantum systems. In general, there are presently two approaches to the problem. The first is based on the correlations between photon arrival times, or the related arrival times of optical signals detected by homodyne detection \cite{Gio01,Bah04,Val04,Lam08}. The second approach is based on the internal time evolution of quantum systems \cite{Buz99,Joz00,Chu00,Pre00,Yur02,Krc02,Boi06,Ben11}. Although the latter approach requires an effective suppression of decoherence and is therefore much more challenging to implement, it might be of greater fundamental interest, since it allows a very general treatment of time in quantum mechanics. Specifically, it can show how the time evolution of quantum systems affects the non-classical correlations between entangled quantum systems.

Initially, it was shown that two-party quantum clock synchronization
protocols can be used for efficient clock synchronization by using
the enhanced sensitivity of bipartite entangled states to small time
differences between the measurements performed by the two parties
\cite{Joz00}. Later, Krco and Paul \cite{Krc02} extended this idea
to a multi party version, where a W-state was used to simultaneously
provide bipartite entanglement between a central clock and several
other parties. However, the bipartite entanglement obtained from the
W-state decreases rapidly with an increase in the number of clocks.
Ben-Av and Exman \cite{Ben11} pointed out that this is a weakness of
the W-state that can be overcome by using other Dicke states
instead. Specifically, they showed that the optimal bipartite
entanglement for this kind of protocol is obtained by using the
symmetric Dicke states, where half of the qubits are in the 0 state
and half are in the 1 state.

Interestingly, none of these protocols uses the specific properties
of multipartite entanglement. The reason for this may be that it is
a straightforward matter to measure and evaluate the correlations
between two parties, while genuine multi-partite entanglement is
characterized by more complicated correlations that involve the
measurement results observed at all locations. Here, we consider the
question of whether this different type of entanglement could be
used for clock synchronization by constructing a protocol that
accesses the maximal entanglement of GHZ-type states through an
appropriate combination of measurement and communication between the
parties. The result should be significant both for determining the
limits of multi party clock synchronization and for our general
understanding of time in multi-partite entanglement.

As we discuss below, a protocol using GHZ-type states requires a specific division of the parties
into groups during each distribution of the entangled qubits, since
there is no GHZ-type state that is both an energy eigenstate and
symmetric in all parties. The logic of the two party protocol can
then be applied to the collective information of all measurements in
the two groups. By selecting an appropriate set of divisions, it is
thus possible to use the complete entanglement of the GHZ-state to
efficiently synchronize all N clocks simultaneously. In the
following, we introduce the protocol, evaluate its efficiency and compare it with the efficiencies obtained from bipartite entanglement in the protocols based on parallel two party synchronization \cite{Joz00} and on the use of symmetric Dicke states \cite{Ben11}.

\section{State preparation and distribution}

Here, we consider a type of clock synchronization protocol where entangled qubits are distributed to various parties holding the individual clocks \cite{Buz99,Joz00,Chu00,Pre00,Yur02,Krc02,Boi06,Ben11}. Each quantum system is described as a two-level spin precessing around the z-axis at a fixed frequency $\omega$. For experimental realizations, it would be important to keep decoherence effects to a minimum, e.g. by using nuclear spins precessing in the intrinsic field of a molecule or crystal \cite{Kan98}. However, it should be kept in mind that the present research is not motivated by such technical considerations, but by an interest in the fundamental nature of quantum clocks, as introduced in the semininal work by Buzek et al. \cite{Buz99}. In this spirit, we assume that after state preparation, the evolution of the internal spin state only depends on the passage of time. The problem of clock synchronization can then be reduced to the problem of identifying the time differences between time-dependent measurements performed on the different clocks.

Since clock synchronization should not depend on a knowledge of the
time needed for state distribution, the multipartite entangled
states used should be energy eigenstates. It is therefore not
possible to use GHZ states that are superpositions of the two
extremal eigenstates of energy, where all qubits are in the same
state of their local energy basis. To obtain an energy eigenstate
without changing the multipartite entanglement, half of the local
energy eigenstates should be flipped by appropriate local unitary
transformations. If the qubits are arranged so that the first half
of the qubits is unflipped and the second half of the qubits is
flipped, this $N$-partite entangled energy eigenstate can be given
in the energy basis as
\begin{equation}\label{maximal entanglement}
\mid  \Psi_{N}\rangle=\frac{1}{\sqrt{2}}\left(\mid 0\rangle^{\otimes
\frac{N}{2}}\mid 1\rangle^{\otimes \frac{N}{2}}+\mid
1\rangle^{\otimes \frac{N}{2}}\mid 0\rangle^{\otimes
\frac{N}{2}}\right).
\end{equation}
Here and in the following, we assume an even number of parties $N$.
The states $\mid 0\rangle$ and $\mid 1\rangle$ are local energy
eigenstates with energies $0$ and $\hbar \omega$, respectively.

The state given by Eq.(\ref{maximal entanglement}) divides the
qubits into two groups. To ensure clock synchronization between all
parties, it is necessary that no two parties are always members of
the same group. This is achieved by distributing the qubits in
different ways, so that each party sometimes receives a qubit from
the unflipped group, and sometimes receives a qubit from the flipped
group. To describe each distribution, we define a sequence $\{ f_{i}
\}$, where $i=1,...,N$. If the qubit of the $i-th$ clock owner is a
flipped qubit, $f_{i}=1$, if not, $f_{i}=0$. Since the numbers of
flipped and unflipped qubits are equal, the number of possible
distributions $\{ f_{i} \}$ is given by the binomial coefficient
$N!/((N/2)!(N/2)!)$. In the most simple version of the protocol, the
division into groups can be decided randomly in each run, with equal
probabilities for each distribution $\{ f_{i} \}$.

\section{Measurement and clock synchronization}

After the distribution of the qubits to the locations of the
different clocks, each of the parties measures a time dependent
observable $\hat{X}(t)$ on its qubit when their local clock points
to a specific time. The observable measured at a time $t$ can be
written as
\begin{equation} \label{time evolution operator}
\hat{X}(t) = \exp(-i \omega t) \mid 0 \rangle \langle 1 \mid +
\exp(i \omega t) \mid 1 \rangle \langle 0 \mid.
\end{equation}
The eigenvalues of the measurement outcomes are $\pm 1$. The eigenstates corresponding to the measurement outcomes are equal superpositions of $\mid 0 \rangle$ and $\mid 1 \rangle$, where the phase now depends on the time at which the measurement is performed. As a result, this measurement achieves the maximal time sensitivity for local qubit measurements.

The time sensitivity of the maximal multi-partite entanglement of the GHZ-type energy eigenstate given in Eq.(\ref{maximal entanglement}) originates from the coherence between the components $\mid 0\rangle^{\otimes
\frac{N}{2}}\mid 1\rangle^{\otimes \frac{N}{2}}$ and $\mid 1\rangle^{\otimes\frac{N}{2}}\mid 0\rangle^{\otimes \frac{N}{2}}$. This coherence, which represents the full multi-partite entanglement of the state, changes the probability of the collective measurement outcome depending on the product of the coherences between the $\mid 0 \rangle$ and $\mid 1 \rangle$ components in the eigenstates representing the local measurement outcomes. As a result, the time sensitivity of multi-partite entanglement can be represented by the espectation value for the product of all outcomes, $\hat{X}^{\otimes N}$. If
the actual measurement times of the parties are given by $\{t_{1},
t_{2},..., t_{N}\}$  the expectation value of this product is
\begin{equation} \label{average number}
\langle\hat{X}^{\otimes
N}\rangle=\cos\left(\sum_{i}^{N}(-1)^{f_{i}}\; \omega t_{i}\right).
\end{equation}
Since the local measurements represent the maximal time sensitivity for the local qubits, and since the coherence that characterizes multi-parite entanglement of the GHZ-type can only be observed in the product of all local measurement outcomes, we can conclude that the time dependence shown in Eq.(\ref{average number}) is the strongest dependence on the measurement outcomes $t_i$ that can be achieved with GHZ-type states and local measurements. A protocol that makes use of the time dependent correlations between all of the $\hat{X}$ measurements therefore accesses the full power of maximal multipartite entanglement for clock synchronization.

\begin{figure}
[ht]
\begin{center}
\includegraphics[width=0.2\textwidth]{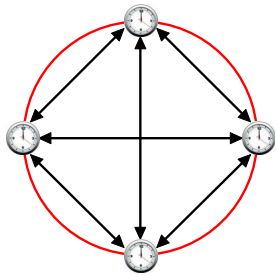} \caption{\label{fig1}
\footnotesize{Illustration of clock synchronization with maximal
multipartite entanglement. The circle indicates the multipartite
entanglement, the arrows indicate classical communication between
the parties. Communication between all parties is necessary to make
the high sensitivity of multipartite entanglement available for
clock synchronization.}}
\end{center}
\end{figure}

To access the time sensitivity of the GHZ-type state, all parties must share their measurement results and determine the product of all outcomes. Effectively, the $N$ parties cooperate to measure a single
$N$-particle interference fringe that is sensitive to the collective
phase given by $\omega$ times the difference between all measurement
times of the unflipped qubits ($f_i=0$) and the measurement times of
all flipped qubits ($f_i=1$). Significantly, the use of maximal multi-partite entanglement for clock synchronization critically depends on simultaneous access to all measurement outcomes. It therefore
requires classical communication between all the parties, as indicated
in Fig. \ref{fig1}. The full sensitivity of maximal multipartite entanglement only becomes available for use in the clock synchronization process, if all of the parties cooperate.

In the final step of the synchronization protocol, the clock owners
have to estimate the time differences between their respective local
clocks and a standard time. Since the present protocol is fully
cooperative, with complete symmetry between all parties, it seems
natural to define this standard time as the average of all clock
times. In the following, we therefore discuss the synchronization to
average time. To achieve synchronization to a standard clock, one
can simply change the adjustments of the time, so that instead of
changing his or her own time, the owner of the standard clock makes
everyone else subtract his or her time adjustment from theirs.

To ensure that all parties are treated equally, it is possible to
use a random distribution of qubits, so that every distribution $\{
f_{i} \}$ of flipped and unflipped qubits is equally likely. To keep
track of the different distributions, we assign an index $j$ to
each, so that the elements of each sequence are given by $f_i(j)$.
The total time difference that defines the phase shift in the
multipartite interference fringe observed in the $\hat{X}^{\otimes
N}$ measurement of the distribution with index $j$ is then
\begin{equation}
T_j = \sum_{i=1}^{N} (-1)^{f_i(j)} t_i.
\end{equation}
The time differences $T_j$ can be estimated from the outcome
statistics of the measurements with an accuracy of $\delta T_j =
1/(\omega \sqrt{k_j})$, where $k_j$ is the number of times that the
distribution $j$ is received and measured.

After a sufficiently large number of measurements, all parties have the same estimates for all possible time differences $T_j$. However, the implications of each $T_j$ are different for each party. Specifically, each party $i$ can obtain the difference of times $T_j$ with $f_i(j)=0$ and the times $T_j$ with $f_i(j)=1$,
\begin{equation}
\sum_{j}(-1)^{f_i(j)} T_j = \sum_j \sum_k (-1)^{f_i(j)}(-1)^{f_k(j)}
t_k
\end{equation}
For $k=i$, the coefficient in the sum is always $+1$, so that the
time $t_i$ of the local clock always enters into the sum with a
positive value. Since all the other times enter into the sum
equally, and since the number of $+1$ coefficients and $-1$
coefficients is exactly equal, the result can be expressed in terms
of the difference between the time $t_i$ and the average $\langle t
\rangle_{k\neq i}$ of all times other than $t_i$,
\begin{equation}
\sum_{j}(-1)^{f_i(j)} T_j = \frac{N!}{(N/2)! (N/2)!}(t_i-\langle t \rangle_{k\neq
i}).
\end{equation}
The average of all times $\langle t \rangle$ is obtained by the
weighted average of $t_i$ and $N-1$ times $\langle t \rangle_{k\neq
i}$. Hence, the difference between the local time $t_i$ and the
average time $\langle t \rangle$ can be given by,
\begin{equation}
t_i-\langle t \rangle = \left(\frac{N-1}{N}\right) \sum_j
(-1)^{f_i(j)} \; \frac{(N/2)! (N/2)!}{N!} \; T_j.
\end{equation}
After this value is determined by each party, it can be subtracted
from each local clock time $t_i$ to adjust the clock times so that
they correspond to the average time $\langle t \rangle$.

\section{Precision of adjustment times}

To determine the efficiency of a clock synchronization protocol, it
is necessary to evaluate the precision with which the parties can
estimate the adjustment time $t_i-\langle t \rangle$. In general,
this precision is limited by the statistical variance of the
measurement results. As mentioned above, the estimation errors for
the time differences $T_j$ are given by $\delta T_j = 1/(\omega
\sqrt{k_j})$, where $k_j$ is the number of times that the
distribution $j$ was measured. Since the adjustment times
$t_i-\langle t \rangle$ are linear functions of the $T_j$, it is
sufficient to find the sum of the quadratic errors with the
appropriate coefficients to obtain the adjustment errors
\begin{equation}
\delta t_i^2 = \sum_j \left(\frac{N-1}{N}\right)^2 \left(
\frac{(N/2)! (N/2)!}{N!}\right)^2 \delta T_j^2.
\end{equation}
If each distribution $j$ is measured an equal number of times, $k_j$
can be expressed as the total number of measurements $k$ divided by
the number of possible distributions $N!/((N/2)!(N/2)!)$. Likewise,
the sum over $j$ reduces to a simple multiplication with the number
of possibilities. In the end, the estimation error for each
adjustment time is given by
\begin{equation}
\delta t_i^2 = \left(\frac{N-1}{N}\right)^2  \frac{1}{\omega^2 k}.
\end{equation}
In the limit of large $N$, this error is simply $\delta t_i =
1/(\omega \sqrt{k})$, independent of the number of parties
participating in the clock synchronization. This means that the
maximally multipartite entangled states can be used to synchronize
$N$ clocks in parallel, without any loss of precision when
additional parties are added.

\section{Comparison with parallel distribution of bipartite entanglement}

Since the parallel synchronization of $N$ clocks can also be
achieved by performing separate synchronizations of $N-1$ clocks
with the same standard clock, it is not immediately clear whether
multipartite entanglement has any advantages over multiple bipartite
entangled states. In the following, we therefore analyze the
efficiency of multipartite clock synchronization using the initial
proposal for quantum clock synchronization between two parties
\cite{Joz00}. The complete multi party protocol is illustrated in
Fig.2. There are $N$ spatially separated unsynchronized clocks, one
of which is the standard clock. The remaining $N-1$ clock owners
synchronize their clocks with this standard clock using bipartite
entanglement and classical communication. For this purpose, the
owner of the central clock must share $N-1$ maximally entangled two
qubit states with all of the other parties for each measurement.
Effectively, each step of the protocol uses a $2(N-1)$ qubit state
given by
\begin{equation}
\label{2qubit} \mid  \mbox{parallel}
\rangle=\left(\frac{1}{\sqrt{2}}\right)^{N-1}  \left(\mid
01\rangle+\mid 10\rangle\right)^{\otimes(N-1)}.
\end{equation}
Here, every second qubit is held by the owner of the central clock.
At a predetermined time $t$, the owner of the central clock measures
the value of $\hat{X}_{ci}(t)$ on all of her $N-1$ qubits $i$, where
$i$ is the index of the party that holds the qubit entangled with
the qubit $ci$. Likewise, the other parties measure the value of
$\hat{X}_{pi}(t_i)$ on their individual qubits according to their
local times $t_i$. The central clock then communicates each result
of $\hat{X}_{ci}(t)$ to the party concerned. After a sufficiently
large number of measurements $k$, the owner of clock $i$ can then
determine the expectation value of the product,
\begin{equation} \label{2Xcorr}
\langle\hat{X}_{pi}\otimes\hat{X}_{ci}\rangle=\cos(\omega(t_{i}-t)).
\end{equation}
The clock owners can then determine $t_i-t$ directly and adjust
their clocks accordingly.

\begin{figure}
[ht]
\begin{center}
\includegraphics[width=0.2\textwidth]{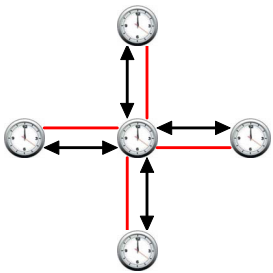}
\caption{\label{fig2} \footnotesize{Illustration of clock
synchronization using parallel distribution of bipartite entangled
states. The arrows indicate classical communication and the lines
indicate shared bipartite entanglement. Each of the outer $N-1$
clocks is synchronized separately with the central clock using
separate sets of entangled qubits.
}}%
\end{center}
\end{figure}

The efficiency of clock synchronization can be evaluated by
considering the estimation error $\delta t_i$ for each estimate of
adjustment time $t_i-t$. For $k$ measurements, this error is given
by $\delta t_i=1/(\omega \sqrt{k})$. Thus the precision of the time estimates in this protocol is exactly equal
to the result for the protocol using multipartite entanglement.
However, the parallel distribution of bipartite entanglement
requires $2(N-1)$ qubits for each measurement, as compared to only
$N$ qubits for the multipartite entangled protocol. In terms of the
required number of qubits, the use of multipartite entangled states
can thus increase the efficiency by a factor of two. Effectively, the main effect of
multipartite entanglement seems to be that the need for multiple
reference qubits held by the owner of the central clock is removed by allowing
the parties to use the $N-1$ qubits of all the other parties as a
collective reference instead.

\section{Comparison with the symmetric Dicke state protocol}

\begin{figure}
[ht]
\begin{center}
\includegraphics[width=0.2\textwidth]{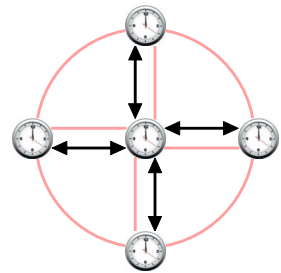}
\caption{\label{fig3} \footnotesize{Illustration of clock
synchronization using an N-qubit symmetric Dicke state. The arrows
represent classical communication, the lines indicate the
entanglement between the qubits. Since each qubit is entangled with
all the other qubits, the owner of the central clock only needs to
hold a single qubit per measurement.}}
\end{center}
\end{figure}

Previous multi party clock synchronization protocols were based on
parallel clock synchronization using the bipartite entanglement
available from W-states \cite{Krc02} or from symmetric Dicke states
\cite{Ben11}. In particular, Ben-Av and Exman showed that the
symmetric Dicke state is optimal in the sense that it maximizes the
bipartite entanglement between the single qubit held by the owner of
the central clock and the qubits held by all of the other parties
\cite{Ben11}. Their protocol is illustrated in Fig. \ref{fig3}. It
uses the same measurement and communication procedure as in the
parallel distribution of entanglement, but with only a single qubit
at the central clock for a total number of $N$ qubits per
measurement - the same number as our GHZ state protocol, and almost
half the number of qubits used in the parallel distribution
protocol.

The symmetric Dicke states is an equal superposition of all energy
eigenstates with half of the qubits in the $\mid 0 \rangle$ state
and half in the $\mid 1 \rangle$ state,
\begin{eqnarray} \label{Symmetric Dicke state}
\mid \mbox{Dicke}(N)\rangle=\frac{\frac{N}{2}!}{\sqrt{N!}}(\mid
11...10...00\rangle\nonumber\\+\mid 11...01...00\rangle+...+\mid
00...01...11\rangle).
\end{eqnarray}
The bosonic symmetry of the state means that the qubits tend to be
found in the same superposition states of $\mid 0 \rangle$ and $\mid
1 \rangle$, resulting in positive correlations between the values of
$\hat{X}(t_i)$ obtained by the different parties at the same time
$t_i$. Specifically, the correlation between the measurement at the
central clock and the measurement at clock $i$ is given by
\begin{equation} \label{Dicke2Xcorr}
\langle\hat{X}_{pi}\otimes\hat{X}_{ci}\rangle= \frac{N}{2(N-1)}
\cos(\omega(t_{i}-t)).
\end{equation}
At the maximal time derivative of the expectation value, the error
in the adjustment time $t_{i}-t$ for $k$ measurements is given by
$\delta t_i=2(N-1)/(N \omega \sqrt{k})$. In the limit of large $N$,
this error is equal to twice the error of our GHZ state protocol and
the parallel distribution protocol. Hence, this protocol requires
four times as many qubits to achieve the same accuracy as the GHZ
state protocol, and twice as many qubits as the parallel
distribution protocol. The reduction in qubit number over parallel
distribution of bipartite states is therefore more than offset by
the loss of sensitivity in each individual measurement due to the
reduction in the available bipartite entanglement.

\section{Qubit efficiencies for clock synchronization}

We can now summarize our results in terms of the accuracy of clock
synchronization achieved with a given number of qubits. Since the
timescale is defined by the resonant frequency $\omega$ of the qubit
dynamics, it is convenient to define the relative accuracy as
$1/(\omega \delta t_i)^2$. For the GHZ-type multipartite
entanglement, the accuracy of $k$ measurements using $Q= k N$ qubits
is then given by
\begin{equation}
\left.\frac{1}{(\omega \delta t_i)^2}\right|_{\mbox{GHZ}} =
\left(\frac{N}{N-1} \right)^2 \frac{Q}{N}.
\end{equation}
For high $N$, the accuracy is equal to the number of qubits per
party, so the accuracy of the multi party protocol scales linearly
with the ratio of qubits and parties, $Q/N$.

Significantly, the straightforward extension of the bipartite
protocol by parallel distribution of entangled qubit pairs performs
only half as well. Specifically, the accuracy of $k$ measurements
using $Q= 2 k (N-1)$ qubits is
\begin{equation}
\left.\frac{1}{(\omega \delta t_i)^2}\right|_{\mbox{pairs}} =
\frac{1}{2} \left(\frac{N}{N-1}\right) \; \frac{Q}{N}.
\end{equation}
The need for extra reference qubits held by the owner of the central
clock therefore rapidly reduces the efficiency of each qubit to half
the value achieved by the protocol using maximal multipartite
entanglement.

Finally, the protocol using the simultaneous bipartite entanglement
between a single central qubit and $N-1$ others achieves a
sensitivity reduced by a factor of $N/(2 (N-1))^2$ due to the
reduction in bipartite entanglement associated with the increase in
entangled partners for each qubit. The accuracy of $k$ measurements
using $Q=k N$ qubits is therefore
\begin{equation}
\left.\frac{1}{(\omega \delta t_i)^2}\right|_{\mbox{Dicke}} =
\frac{1}{4} \left(\frac{N}{N-1}\right)^2 \; \frac{Q}{N}.
\end{equation}
In the limit of high $N$, this is a reduction to one quarter of the
GHZ-type protocol, twice as much as the reduction in accuracy due to
the additional reference qubits in the parallel distribution
protocol.

\section{Conclusions}
We have shown how the maximal $N$-partite entanglement of GHZ-type
stated can be used for multi party clock synchronization by randomly
dividing the parties into two groups during each run and sharing the
measurement results with all other parties to determine the
adjustments necessary to set each local clock to the average time of
all clocks. The accuracy of clock synchronization corresponds to the
accuracy achieved by $N-1$ bipartite protocols in parallel, but the
number of qubits used is reduced by half. Oppositely, the previously
proposed use of symmetric Dicke states uses the same number of
qubits, but the accuracy is only one quarter due to the reduced
amount of bipartite entanglement.

Our results show that the full power of maximal multipartite
entanglement can improve the performance of clock synchronization by reducing the number of qubits needed to achieve a given accuracy by a factor of two when compared to the most efficient use of bipartite entanglement. Although this is clearly an improvement, it is much less than the improvements of sensitivity when a single parameter is estimated using multi-partite entangled probes. The reason for this limited improvement is that $N$ different clock times must be estimated from the same measurement result, leading to a reduction of precision that exactly compensates the gain caused by the increased sensitivity to an average shift in time. From the viewpoint of an individual clock owner, multi-partite entanglement merely replaces the single central clock used in parallel clock synchoronization with bipartite entangled states with the collective of all other $N-1$ clock owners. Overall, the efficiency increases by a factor of two, because the simultaneous role of all clock owners as participants and as reference overcomes the need for additional reference qubits. In this sense, multi-partite entanglement simply represents the simultaneous availability of quantum correlations to all parties, without any increase to the individual time sensitivities.

From a practical viewpoint, the use of multi-patite entanglement may be difficult, since the loss of a single qubit will completely destroy the essential quantum coherence of the state. To obtain results close to the ones described here, the probability of local losses or dephasing errors must be kept far below $1/N$. Oppositely, this sensitivity to decoherence also highlights the cooperative nature of the protocol: if only a single party sends wrong information, the synchronization of the clocks becomes impossible. Thus, clock synchronization with maximal multipartite entanglement also highlights the cooperative nature of multi-party quantum protocols.

In conclusion, the analysis presented here shows how the full power of maximal multipartite
entanglement can be used to improve the performance of clock synchronization if all of the parties involved cooperate to share their measurement information. The results may provide interesting insights, both into the role of entanglement in clock synchronization protocols, and into the fundamental nature of time-dependent quantum correlations.

\section*{Acknowledgment}
Part of this work has been supported by the Grant-in-Aid program of
the Japanese Society for the Promotion of Science, JSPS.

\vspace{-0.5cm}


\end{document}